\documentstyle[12pt]{article}

\newcommand{\de}{\partial}
\newcommand{\be}{\begin{equation}}
\newcommand{\ee}{\end{equation}}
\newcommand{\bea}{\begin{eqnarray}}
\newcommand{\eea}{\end{eqnarray}}
\newcommand{\bd}{\begin{displaymath}}
\newcommand{\ed}{\end{displaymath}}

\def\fey{{\big / \kern-.80em D}}
\font\mybb=msbm10 at 12pt

\def\bb#1{\hbox{\mybb#1}}

\def\zet{{\bb{Z}}}
\def\real{{\bb{R}}}

\def\R4{\real^4}
\def\ie{{\it i.e.}}

\def\unita{{1 \kern-.30em 1}}

\begin{document}
\thispagestyle{empty}
\vskip 1.5cm
\begin{flushright}
ROM2F--96--53 \\
{\tt hep-lat/9610008}
\end{flushright}
\centerline{\large \bf A WILSON--MAJORANA REGULARIZATION} 
\centerline{\large \bf FOR LATTICE CHIRAL GAUGE THEORIES}
\vspace{2.2cm}
\centerline{\sc GABRIELE TRAVAGLINI}
\vskip 0.1cm
\centerline{\sl 
Dipartimento  di Fisica, Universit\`a di Roma ``La Sapienza"}
\centerline{\sl I.N.F.N. \ -- \  Sezione di Roma II, }
\centerline{\sl Via della Ricerca Scientifica, \ \ 00133 \ Roma, \ \ ITALY}
\vskip 2.2cm
\centerline{\bf ABSTRACT}
{We discuss the regularization  of  chiral gauge theories
on the  lattice introducing only physical degrees of freedom.
This is obtained by writing the Wilson term in a Majorana form, 
at the expense of the 
\mbox{$U(1)$} symmetry related to fermion number conservation. 
The idea of restoring chiral invariance in the continuum 
by introducing a properly chosen set of counterterms to be added to 
the tree level action is checked against 
one--loop perturbative calculations.}
\vskip .2in
\centerline{Nuclear Physics B (in press)}
\vskip .1in

\newpage
\setcounter{page}{1}
\setcounter{equation}{0}
\setcounter{section}{0}
\section{Introduction}

\indent 


The problem of finding a nonperturbative formulation of a 
chiral gauge theory is  still an active field of investigation
\cite{Borrelli,Kaplan,Maianiw,Neuberger,fro,Hooft,Hernandez,Alonso,Bietenholz,
Hsu,Bodwin,Shamir,SG,nuovo}.   
The root of this difficulty lies in the  impossibility of finding 
a regularization procedure which preserves
the chiral symmetry of the classical action functional.
\mbox{QED} gives us  a first example of this conflict 
with the Adler--Bell--Jackiw anomaly \cite{Adler,Bell}, 
which spoils the conservation  of the axial current. 
The anomaly term happily accounts for the measured 
$\pi^{0} \rightarrow 2 \gamma$ decay rate, 
which otherwise would be suppressed
according to the Sutherland--Veltman theorem \cite{IZ}.
In supersymmetric  gauge theories a number of correlators 
identically vanishing in perturbation theory 
are in fact nonzero thanks  to the  existence 
of topologically nontrivial gauge 
field configurations  with important  consequences, as 
spontaneous supersymmetry breaking in certain chiral theories
\cite{AR0,afdisei,AR}.
But even more striking is what happens in  the Weinberg--Salam model, 
because of the severe physical constraints imposed by
the cancellation of quark against lepton anomalies, in order to ensure 
gauge invariance  of the theory,  and hence renormalizability \cite{Gross}.

The key point is that it is impossible to quantize chiral fermions coupled to 
gauge fields  in a way that simultaneously preserves the gauge symmetry and the 
validity of the usual  axioms of Quantum  Field Theory. 
In the context of the lattice 
regularization of a chiral gauge theory with gauge group $SU(2)$,
we shall study  the possibility of restoring  chiral symmetry
when  the cutoff $\Lambda$ (the inverse lattice spacing, $a^{-1}$) goes to 
infinity.
In our formulation, discussed in section 2, we shall not need to introduce any 
unphysical degrees of freedom, as suggested in 
\cite{Maianiw,Maiani,Testa}. In section 3, we will compare the 
symmetries of the classical action with those of the regularized one. 
In particular, in section 4, we briefly discuss on the 
anomaly structure of the global $U(1)$ currents and on the related question 
of fermion number nonconservation. 
Following the approach of \cite{Borrelli}, we shall replace 
{\em from  the start}
the asymmetrical regularized action $S$ with an  
(asymmetrical) functional \mbox{$S + S_{c.t.}$}, where 
$S_{c.t.}$  contains all the 
interactions which are 
not  invariant under the symmetries broken by the 
regularization (but of course still invariant under the preserved ones).
We shall choose these interactions in such a way that their canonical 
dimension is less than or equal to four, to keep  renormalizability. 
Moving from the  observation that the formal continuum theory  
is BRST--invariant, we shall  try to satisfy the related 
Slavnov--Taylor identities in the regularized theory
with action $S+S_{c.t.}$
{\em in the limit of infinite cutoff}, that is up to terms vanishing when 
\mbox{$a \rightarrow 0$};
this  way we fix the coefficients  of the interactions
in $S_{c.t.}$.
The  program we have just outlined is checked in section 5 by performing 
a few one--loop computations. In the last section we finally 
draw some conclusions.

\section{Description of the Model}
\setcounter{equation}{0}
In this  model we shall consider a fermionic undotted Weyl 
isodoublet\footnote{We are 
using van der Waerden's notation for Weyl spinors \cite{Waerden}.}
\mbox{$\chi^{\alpha}_{A} (n)$}, 
$A = 1,2$, defined on lattice sites which 
transforms according to the fundamental representation of the group
\mbox{$SU(2)$},
\mbox{$\chi^{\alpha} (n) \ \longrightarrow \ \Omega (n) \chi^{\alpha} (n)$}, 
\mbox{$\Omega (n) \in SU(2)$}. Its dotted partner
\mbox{$\bar{\chi}_{\dot{\alpha}, A}(n)$}  transforms as
\mbox{$\bar{\chi}_{\dot{\alpha}}(n) \longrightarrow  
\bar{\chi}_{\dot{\alpha}}(n)  \Omega^{\dagger}(n) $}, and is 
an independent variable from the point of view 
of (Euclidean) path integration.
We focus  our analysis on a theory based on the  simple gauge group 
\mbox{$SU(2)$}, whose  representations are 
real, because we want to keep the calculations as simple as possible;  
\ie\  we avoid the introduction of more fermion matter fields, which would 
otherwise be necessary to cancel local anomalies, as in the Standard Model.
This restriction is merely of a technical  nature. Of course one can consider
different theories with fermions making up anomaly--free representations; 
the reader can find
a discussion concerning the case of the group $SU(5)$ with one 
generation of ${\bf 5^{\ast}} + {\bf 10}$ 
Weyl fermions in \cite{Maiani}. 
It is well known that an $SU(2)$ gauge theory with one Weyl fermion 
in the fundamental representation is plagued by a
global anomaly \cite{Witten};  nevertheless this 
will not affect  our calculations, because they are 
performed around the trivial vacuum configuration.

Let then \mbox{$U_{\mu}(n)=\exp  \Bigl[ iag A_{\mu} (n + \mu / 2 ) \Bigr] $}
be a link variable belonging to \mbox{$SU(2)$}\footnote{Our 
gauge connection is hermitian and it is defined  as
$A_{\mu} \equiv A_{\mu}^{a} \sigma^a / 2$, where 
the $\sigma^a$  are the usual Pauli matrices.}
and transforming as  
\mbox{$U_{\mu}(n) \longrightarrow 
\Omega(n) U_{\mu}(n)\Omega^{\dagger}(n + \mu) $}
under the gauge group.
The action functional  \mbox{$S_{T} + S_{YM} $}, where 
\be
S_{T}=\frac{1}{2a}\sum_{n, \mu} a^4 \ \bigg[ \bar{\chi}^{\dot{\alpha}}(n)
\overline{\sigma}_{\dot{\alpha}\beta , \mu}U_{\mu}(n)\chi^{\beta}(n + \mu)- 
\bar{\chi}^{\dot{\alpha}}(n + \mu)U_{\mu}^{\dagger}(n)
\overline{\sigma}_{\dot{\alpha}\beta , \mu}\chi^{\beta}(n)  \bigg] \ ,
\ee
\be
S_{YM}=-\frac{1}{2g^2}\sum_{n,\mu ,\nu} \ 
\mbox{{\rm Tr}} \left[ U_{\mu}(n)U_{\nu}
(n + \mu )U_{\mu}^{\dagger}(n + \nu ) U_{\nu}^{\dagger}(n)  - 1 \right] \ + \ 
\mbox{{\rm h.c.}} \ ,
\ee
is then invariant with respect to local $SU(2)$
transformations.\footnote{A brief excursus on 
$\bar{\sigma}$ matrices can be found in the 
appendix.}

Unfortunately, the theory defined by \mbox{$S_{T} + S_{YM} $}
does not correctly reproduce the continuum limit of a chiral gauge theory, due 
to the well--known fermion species doubling.
The strategy proposed by Wilson, to circumvent this problem, consists in adding 
to the action a term chosen in such a way that only one fermion species 
survives 
in the continuum limit, \mbox{$a \rightarrow 0$}, having given 
the spurious states a mass of \mbox{order $a^{-1}$}
\cite{Wil}. 
Using Weyl spinors it is  possible to choose this term in a way in which  
only left--handed fermions are involved, that is without introducing any 
fictitious degree of freedom
\cite{Maianiw,Maiani,Testa}.
This interaction (that we call the Wilson--Majorana term) is 
\bea
\label{Wilson}
S_{W}& = &
-\frac{r}{a}\sum_{n,\mu}a^4 \biggl\{ \bar{\chi}^{\dot{\alpha}}(n) \biggl(
\bar{\chi}^{\dot{\beta}}(n+\mu) - \bar{\chi}^{\dot{\beta}}(n) \biggr)
\epsilon_{\dot{\alpha}\dot{\beta}} +  
\nonumber \\
& & \biggl( \chi^{\alpha}(n+\mu) - 
\chi^{\alpha}(n) \biggr) \chi^{\beta}(n)\epsilon_{\beta\alpha} \biggr\} \ ,
\eea
where \mbox{$ 0 <  r \leq 1$},
and   \mbox{(\ref{Wilson})} is  the discretized version of
\bea
-\frac{1}{2} (ra)\int\! d^{4}x\ \Bigl( \bar{\chi}^{\dot{\alpha}} \Box 
\overline{\chi}^{\dot{\beta}} \epsilon_{\dot{\alpha}\dot{\beta}} + 
\chi^{\alpha}\Box\chi^{\beta}\epsilon_{\beta\alpha} \Bigr)
\ ,
\eea
which is an irrelevant coupling, formally vanishing when the cutoff is 
removed.
We can immediately check that this term gives a mass of order
$r a^{-1}$ to the unwanted fermionic degrees of freedom 
by looking at  the two point Green functions of the free theory,
\be
\label{Leonard}
P^{\dot\alpha \beta}_{AB} (n) \equiv
\left\langle \chi^{\beta}_{A}(n) 
\bar{\chi}^{\dot{\alpha}}_{B}(0) \right\rangle
 = \int_{-\frac{\pi}{a}}^{\frac{\pi}{a}}\!\frac{d^{4}p }{(2\pi)^4}
 e^{ipna}\  \frac{\Bigl( -
ia\sum_{\mu}\overline{\sigma}^{\dot{\alpha}\beta}_{\mu}
\sin p_{\mu}a\Bigr)}{\triangle (pa)} \delta_{AB} \ ,
\ee
\be
\label{Friederich}
P^{\alpha \beta}_{AB} (n)\equiv
\left\langle \chi^{\alpha}_{A}(n) \chi^{\beta}_{B}(0) \right\rangle = 
\int_{-\frac{\pi}{a}}^{\frac{\pi}{a}}\!\frac{d^{4}p }{(2\pi)^4} e^{ipna}
\frac{aW(pa)}{\triangle (pa)}\epsilon^{\alpha\beta} \delta_{AB}  \ ,
\ee
\be
\label{Nikolaus}
P_{\dot \alpha \dot \beta , AB} (n)\equiv
\left\langle \bar{\chi}_{\dot{\alpha}, A}(n)
\bar{\chi}_{\dot{\beta}, B}(0) 
\right\rangle =
\int_{-\frac{\pi}{a}}^{\frac{\pi}{a}}\!\frac{d^{4}p }{(2\pi)^4}
 e^{ipna}\ \frac{aW(pa)}{\triangle (pa)}
\epsilon_{\dot{\beta} \dot{\alpha}} \delta_{AB} \ ,
\ee
where
\be
\label{deftrian}
\triangle (p) = D(p) + W^2(p) \ ,
\ee
with 
\be
\label{didipi}
D(p) = \sum_{\mu} \sin ^2 p_{\mu}  \ , 
\ee
\be
\label{defdivu}
W(p) = 2r\sum_{\mu}(1-\cos p_{\mu}) \ .
\ee
The appearance of two nondiagonal propagators in 
\mbox{(\ref{Friederich})}, 
\mbox{(\ref{Nikolaus})}
is the natural consequence of the explicit breaking of 
the \mbox{$U(1)$} symmetry related to fermion number conservation that 
our regularization entails. 
When \mbox{$r \rightarrow 0^{+}$} this symmetry is 
restored, the nondiagonal propagators vanish, 
and we have again \mbox{$2^4$}  different fermions.

To sum up, the complete tree level regularized action of the theory is
\be
\label{action}
S = S_{T} + S_{YM} +  S_{gf} + S_{W} \ \ .
\ee
In the last equation $S_{gf}$ is the sum of  gauge--fixing and ghost terms, 
and reads, in continuum notation, 
\be
S_{gf}= \int\! d^{4}x \ \Bigl[ 
\frac{1}{2\alpha}(\partial_{\mu}A_{\mu}^{a})(\partial_{\nu}A_{\nu}^{a})
\ + \ \bar{c}^{a}\partial_{\mu} \Bigl( D_{\mu} c \Bigr)^{a} \Bigr] \ ,
\ee
where $D_{\mu}^{ab}$ is 
the covariant derivative in the adjoint representation, 
\be
\begin{array}{c c}
D_{\mu}^{ab}= \partial_{\mu}\delta^{ab} + g \ \epsilon^{abc}A_{\mu}^{c}
\ \ \ & \ a\ ,\ b=1, \ 2,\ 3\ \ \ \ .
\end{array}
\ee
In  \cite{Sarno1,Sarno2}, it has been pointed out that in a 
chiral gauge theory the nonsymmetrical ({\it e.g.} 
dimensional) regularization
requires the introduction of nongauge--invariant counterterms 
in the ghost sector, making it
unavoidable to fix the gauge even on the lattice, in contrast to the case 
of a theory with vector--like symmetry 
(where the invariant formulation and the 
compactness of the group make the functional integral well--defined, 
without the need of introducing any supplementary condition).  

Note that we could also add a Majorana mass term to $S$,
\be
\frac{m}{2}\sum_{n} a^4  \left[
\overline{\chi}^{\dot{\alpha}}(n)\overline{\chi}^{\dot{\beta}}(n)
\epsilon_{\dot{\alpha}\dot{\beta}} + 
\chi^{\alpha}(n)\chi^{\beta}(n)\epsilon_{\beta\alpha} \right]   \ ,
\ee
with the sole  effect of  redefining   the function in 
\mbox{(\ref{defdivu})},
\mbox{$W(pa) \longrightarrow W(pa) + ma $}. We shall soon see that 
renormalization generates such a mass term.

\section{Symmetry Properties of the Regularized Action}
\setcounter{equation}{0}
Due to the presence of a gauge--fixing term, 
the action functional $ S-S_{W}=S_T + S_{YM}+S_{gf}$ no longer 
possesses  a gauge symmetry. Conversely, it exhibits a \mbox{BRST} invariance 
\cite{Becchi,Tyutin},
summarized by the infinitesimal transformations
\[
\begin{array}{c c}
\delta\bar{c}^a = \bar{\epsilon} / \alpha g \  
\left( \partial_{\mu} A_{\mu}^{a} \right)    \  \ , & \  \ 
\delta c^{a} = -\bar{\epsilon} / 2 \ \epsilon^{abl}c^{b} c^{l} \ \ ,
\end{array}
\]
\be
\label{trasfbrst}
\delta A_{\mu}^{a} = -\frac{\bar{\epsilon}}{g} \left( D_{\mu} c \right)^{a} \ ,
\ee
\[
\begin{array}{c c}
\delta \chi = i\bar{\epsilon}  c^a T^a \chi \ \ , & \ \ 
\delta \bar{\chi} = i \bar{\epsilon} \bar{\chi} T^a c^a \ \ ,
\end{array}
\]
where $\bar{\epsilon}$ is a Grassmann number, and 
\mbox{$T^{a} = \sigma^{a} / 2$}.
However, $S$ is not invariant under the transformations (\ref{trasfbrst}):
 once we introduce the Wilson--Majorana term,
\mbox{BRST} invariance is lost, as well as the invariance under global  
\mbox{$SU(2)$} and \mbox{$U(1)$} transformations:
in the language of a Wilson--like lattice regularization, this
is the way anomalies come into play.
The breaking of \mbox{BRST} invariance in our model 
is not a consequence of the
form in which we wrote the Wilson term, as stated by the 
Nielsen--Ninomiya theorem:  whatever  strategy we use to remove fermion 
doublers,
chiral  invariance is lost, if we want to preserve the locality of 
the theory \cite{Nielsen1,Nielsen2}.

Let us now move on to the residual symmetry properties of the model.
This question is particularly important here because of the more 
``asymmetrical"
form of our regularization (with respect to the Wilson--Dirac one employed in 
\cite{Borrelli}). This way, we shall still be able to put some restrictions 
on the possible form of the regularized (and unrenormalized) Green functions
\mbox{$G_{\Lambda}(x_1 , \ldots , x_n)$}, thus explaining some of the results 
of perturbation theory that could otherwise seem fortuitous.
We find that two simple discrete symmetries and a continuous one survive 
even for \mbox{$a \neq 0$}. In fact, 
the following three sets of 
transformations leave  the action of the theory invariant:
\begin{itemize}
\item [{\bf 1.}]
\[   
   \begin{array}{c c}
   \chi \longrightarrow e^{i \varphi  T^{2}_{f}}\chi\ \ & \ \  
   \bar{\chi} \longrightarrow \bar{\chi}e^{-i \varphi T^{2}_{f}}  , 
   \end{array} 
\]   
\be
\label{d1}
   A \longrightarrow e^{i \varphi T^{2}_{f}} A e^{-i \varphi T^{2}_{f}} \ ,  
\ee
   \[
   \begin{array}{c c}
   c \longrightarrow e^{i \varphi T^{2}_{adj}} c \ \ & \ \ 
   \bar{c} \longrightarrow \bar{c} e^{-i \varphi T^{2}_{adj}} \ ,
   \end{array} 
   \]

\item [{\bf 2.}]
   \[
   \begin{array}{c c}
   \chi \longrightarrow \sigma^{1} \chi\ \ & \ \  
   \bar{\chi} \longrightarrow \bar{\chi} \sigma^{1}  , 
   \end{array} 
   \]
\be
\label{d2}
   A \longrightarrow e^{i \pi T^{1}_{f}} A e^{-i \pi T^{1}_{f}} \ , 
\ee  
    \[
    \begin{array}{c c}
    c \longrightarrow e^{i \pi T^{1}_{adj}}c \ \ & \ \ 
    \bar{c} \longrightarrow \bar{c} e^{-i \pi T^{1}_{adj}} \ ,
    \end{array} 
    \]

\item [{\bf 3.}]
   \[   
   \begin{array}{c c}
   \chi \longrightarrow \sigma^{3} \chi\ \ & \ \  
   \bar{\chi} \longrightarrow \bar{\chi} \sigma^{3}  \ ,
   \end{array} 
   \]
\be
\label{d3}
   A \longrightarrow e^{i \pi T^{3}_{f}} A e^{-i \pi T^{3}_{f}} \ , 
\ee  
     \[
     \begin{array}{c c}
     c \longrightarrow  e^{i \pi T^{3}_{adj}}c \ \ & \ \ 
     \bar{c} \longrightarrow \bar{c} e^{-i \pi T^{3}_{adj}} \ , 
     \end{array} 
     \]
\end{itemize}
where \mbox{$T^{a}_{r}\ , \ a=1,2,3$} are the generators of the 
symmetry in the representation \mbox{\bf r} (then  
\mbox{$T^{a}_{f}=\sigma^{a} / 2 $}), and $\varphi$ is a constant.
It should be noticed anyway 
that much of this is a 
consequence of the well--known 
property of the generators of \mbox{$SU(2)$} in 
their fundamental representation, 
$
\sigma^2 \sigma^a \sigma^2 = - \sigma^{a \ \ast} 
$.

As we can see, the  first one is a  continuous  global symmetry, whereas the 
last two  involve discrete transformations of the fields.
As a consequence, there will be one conserved Noether current, 
whose expression  is 
\bea
\label{current}
J_{\mu}(n)  & = &  \frac{1}{2}\biggl[ 
\bar{\chi}^{\dot{\alpha}}(n) \overline{\sigma}_{\dot{\alpha} \beta , 
\mu}\frac{\sigma_2}{2}\chi^{\beta} (n + \mu) +
\bar{\chi}^{\dot{\alpha}}(n+ \mu) \overline{\sigma}_{\dot{\alpha} \beta , 
\mu}\frac{\sigma_2}{2}\chi^{\beta} (n) \biggr] \   
\nonumber \\
& & - \ r \biggl[ 
\bar{\chi}^{\dot{\alpha}}(n) \frac{\sigma_2}{2}\bar{\chi}^{\dot{\beta}}
(n+ \mu)  \epsilon_{\dot{\alpha}\dot{\beta}} - 
\chi^{\alpha}(n + \mu)\frac{\sigma_2}{2}\chi^{\beta}(n) \epsilon_{\beta \alpha}
\biggr] \ .
\eea
Since
\[
\bar{\chi}^{\dot{\alpha}}(n) \frac{\sigma_2}{2}\bar{\chi}^{\dot{\beta}}(n)  
\epsilon_{\dot{\alpha}\dot{\beta}} \ = \ 
\chi^{\alpha}(n )\frac{\sigma_2}{2}\chi^{\beta}(n) \epsilon_{\beta \alpha}
\ = \ 0 \ ,
\]
the formal continuum limit of (\ref{current}) is  
\be
J_{\mu}(x) \ = 
\ \bar{\chi}^{\dot{\alpha}}(x) \overline{\sigma}_{\dot{\alpha} \beta , \mu} 
\frac{\sigma_2}{2} \chi^{\beta}(x) \  \ .
\ee
As a simple example, we now discuss the consequences 
of these symmetries on the 
general structure, in color space, 
of the nondiagonal propagator of 
$\chi$ fields \mbox{$S^{\alpha \beta}_{AB} (n) \equiv
\langle \chi^{\alpha}_{A} (n)\chi^{\beta}_{B} (0) \rangle$}. Using the two 
discrete transformations (\ref{d2}), (\ref{d3}),  we obtain that 
\[
S^{\alpha \beta}_{AB} (n) = 
\langle \Bigl(\sigma^{i} \chi^{\alpha}\Bigr)_{A} (n) 
\Bigl( \sigma^{i}\chi^{\beta}\Bigr)_{B} (0) \rangle \ \ \ \ i=1,3\ \ ,
\]
that is 
\be
\begin{array}{c c}
S^{\alpha \beta}_{11} (n) = S^{\alpha \beta}_{22} (n)  \ \  , & \ \ 
S^{\alpha \beta}_{12} (n) = S^{\alpha \beta}_{21} (n) = 0  \ \ .
\end{array}
\ee
Hence \mbox{$S^{\alpha \beta}_{AB} (n) = C^{\alpha\beta}(n) \ \delta_{AB}$}, 
which shows that this propagator is diagonal in color space. 
 
\section{Fermion Number Nonconservation and \newline
Anomalies}
\setcounter{equation}{0}
Let us comment on the breaking of the \mbox{$U(1)$} symmetry
associated with fermion number conservation. 
In the previous formulation of lattice chiral gauge theories
discussed in \cite{Borrelli,Maiani,Testa}
one introduces
\begin{itemize}
\item[{\bf 1.}]
the physical left--handed fermionic degrees 
of freedom, minimally coupled to the gauge fields (in the Dirac notation
of \cite{Borrelli} we call them $\psi_L$), and  
\item[{\bf 2.}]
a set of non--interacting, {\em unphysical}
 right--handed fields, $\psi_R$, to be not confused
with the right--handed physical fields which could possibly be present.
\end{itemize} 
The kinetic term should then be written as
\bea
S_T & = &
\frac{1}{2a}\sum_{n, \mu} a^4 \ \bigg[ \bar{\psi}_L (n)
\gamma_\mu U_{\mu}(n)\psi_L (n + \mu)- 
\bar{\psi}_L (n + \mu) U_{\mu}^{\dagger}(n)
\gamma_{\mu}\psi_L (n)  \bigg] +
\cr
& &
\frac{1}{2a}\sum_{n, \mu} a^4 \ \bigg[ \bar{\psi}_R (n)
\gamma_\mu \psi_R (n + \mu)- 
\bar{\psi}_R (n + \mu) 
\gamma_{\mu}\psi_L (n)  \bigg]\ ,
\eea
which is nothing but the lattice transcription of 
\be
\int d^4x \ 
\biggl[
\bar{\psi}_L
\gamma_\mu (\partial_{\mu} + i g A_{\mu} )\psi_L + 
\bar{\psi}_R 
\gamma_\mu \partial_{\mu} \psi_R 
\ \ .
\bigg]
\ee
The 
$\psi_R$ fields allows one to write a Wilson(--Dirac) term  transforming,
under \mbox{$U(1)$}, like the monomials
$
\bar{\psi}_{L}\psi_{R}
$,
$\bar{\psi}_{R}\psi_{L}
$. 
For example it could be chosen  in such a way that 
its continuum limit is proportional to 
\be
a\int\! d^{4}x \  \biggl[
(\partial_{\mu} \bar{\psi}_{L})( \partial_{\mu} \psi_{R}) + 
(\partial_{\mu} \bar{\psi}_{R})( \partial_{\mu} \psi_{L})  \biggr]
\ \ . 
\ee
The question of fermion number violation in the context of the theory with a 
Wilson term written in a Dirac form has been discussed in \cite{Maianiw,Maiani}.
In this papers  it is argued that fermion number violating correlators can
be obtained  by applying the cluster decomposition theorem to a Green function
like
\be
\label{ca.1}
\left\langle {\cal O}_{\Delta F=+2} (x) {\cal O}_{\Delta F=-2} (y)
\right\rangle
\ \ , 
\ee
where
${\cal O}_{\Delta F=+2}$ (${\cal O}_{\Delta F=-2}$)  describes a transition 
in which the variation $\Delta F$ of the fermion number is +2 (-2).
When $|x-y| \to \infty$,
one gets 
\be
\label{ca.2}
\lim_{|x-y| \to \infty}
\left\langle {\cal O}_{\Delta F=+2} (x) {\cal O}_{\Delta F=-2} (y)
\right\rangle
=
\left\langle {\cal O}_{\Delta F=+2} (x) \right\rangle
\left\langle{\cal O}_{\Delta F=-2} (y)\right\rangle
\ \ ;
\ee
detailed calculations are then required to 
check if the r.h.s. of (\ref{ca.2}) is actually  nonzero.

Of course, in the Majorana form discussed in this paper, clustering arguments 
could still be applied. 
On the other hand, here 
fermion number nonconservation could arise as a consequence of not having 
introduced the unphysical right--handed 
degrees of freedom  which would
allow  us to write a Wilson term  \`{a} la Dirac.
This way we have the  interesting possibility of writing 
non \mbox{$U(1)$--}invariant amplitudes in perturbation theory
for finite lattice spacing, without the need of applying clustering arguments
to Green functions \cite{Maianiw,Maiani,Testa,nuovo}. 
Since the Wilson--Majorana term (\ref{Wilson})
breaks fermion number conservation, 
at the lattice level these amplitudes are nonvanishing. 
Of course   non \mbox{$U(1)$--}invariant correlators 
must vanish in any finite order of
perturbation theory when the cutoff is 
removed; 
 could it happen that,  for some nonperturbative field 
configuration,  some of them have  a nonzero continuum limit ? 
If such a case occurs, we will immediately  be able to identify  
physical  processes in which fermion  number is not conserved
\cite{Hooft1,Hooft2}.
There is however some problem in this approach.
\footnote{We thank the Referee for discussions on this point.}
To begin with, one should require that lattice regularizations reproduce
in the $a\to 0$ limit the Atiyah--Singer index theorem.
It requires that,  if $n_L$ ($n_R$) is the number 
of left--handed (right--handed) zero--modes
of the Dirac operator 
in the representation {\bf r} of the gauge group $G$
in the external field of a gauge connection $A$ of
winding number 
$ k [ A ] $, then
\cite{schwartz,jr,nisc,coleman}
\begin{equation}
n_L - n_R = 2C({\bf r}) k[A]\ \ ;
\end{equation}
here $C({\bf r})$ is the Dynkin index of the representation {\bf r}.
Moreover, 
if $A$ is (anti)self--dual 
$n_R = 0$ ($n_L = 0$), so that  
\begin{equation}
\begin{array}{l l l}
n_L = 2C({\bf r}) k[A]  \ , &\ n_R = 0  \ ,  &\ \  
{\rm if}\  F[A] = \tilde{F}[A] \ \ ,
\\  \\ 
n_R =- 2C({\bf r}) k[A]  \ ,  &\ n_L = 0  \ ,  &  \ \ 
{\rm if}\ F[A] = -\tilde{F}[A] \ \ .
\end{array}
\end{equation}
The index of the {\it Weyl} operator in the representation {\bf r}
in a gauge field of positive winding number  $k[A]$
is then given by $2C({\bf r}) k[A]$. 
However, when the number of fermions per unit volume is
keep finite and the lattice regularized version of the Weyl operator
is a square matrix, the index cannot be nonzero, so that
the index theorem seems to be not correctly reproduced
\cite{na-vra,neuinf}. 

An alternative way to state the problem  is the following.
The number of fermion zero--modes of the (chiral) Weyl operator 
is deeply related to the anomaly of the current associated with fermion number.
The continuum anomaly equation
for the (classically  conserved)
$U(1)$ current for the Weyl fermion 
in the representation {\bf r} of $G$
(in our case $G=SU(2)$ and {\bf r} = {\bf 2})
$J_{\mu}= \bar{\chi}\bar{\sigma}_{\mu}\chi $,
reads   
\begin{equation}
\label{ar.1} 
\partial_{\mu} J_{\mu} = - \frac{1}{32 \pi^2}
(F_{\mu \nu}^{a} \tilde{F}_{\mu \nu}^{a} )  2 C({\bf r}) \ \ ,
\end{equation}
(recall that $C({\bf 2})= 1/2$).
By consequence, a discrete $\zet_{2 C({\bf r})}$ subgroup 
of the anomalous $U(1)$ is preserved
at the quantum level. Since in the present case  
$C({\bf 2})= 1/2$ no such discrete subgroup should survive.
The number of fermion zero--modes is $2 \cdot 1/2\cdot k[A]=k[A]$,
so that, when $k[A]=1$,  one should find
a nonzero vacuum expectation value in the external field of an instanton
for a {\it single}  fermion insertion.
\footnote{In the external field of an instanton, the 
({\it e.g.} continuum) theory has no
Lorentz (as well as conformal) invariance. This is because the instanton
just breaks this symmetry, which is restored after the integration 
over the instanton collective coordinates.}
However, the  regularization here employed,
which is encoded in the Wilson--Majorana term 
(\ref{Wilson}), has a discrete $\zet_2$ symmetry
(which in this case is also a subgroup of the hypercubical 
symmetry of the lattice)
\[   
   \begin{array}{c c}
   \chi \longrightarrow - \chi \ &  \ , \ \  
   \bar{\chi} \longrightarrow - \bar{\chi}\ \ , 
   \end{array} 
\]   
which would imply the absence of fermionic condensates, thus not correctly
reproducing instanton physics \cite{Hooft1,Hooft2}.

This problem is also present when considering the $SU(5)$ gauge theory
with one left--handed fermion, $\chi^{i}$,
in the antifundamental (${\bf 5^{\ast}}$), and  one left--handed fermion,
$\chi_{[ij]}$, in the 2--index antisymmetric representation
({\bf 10}), discussed in \cite{Maiani}. 
There,  one has
$C({\bf 10}) = 3/2$ (and of course
$C({\bf 5^{\ast}} ) = 1/2$). 
The continuum currents 
$J^{\mu}_{\bf 5^{\ast}}$, $J^{\mu}_{\bf 10}$  are then anomalous,
\be
\label{Mozart}
J^{\mu}_{\bf 5^{\ast}}= \bar{\chi}_{i}\bar{\sigma}^{\mu}\chi^{i}
\ \ , \ 
\de_{\mu} J^{\mu}_{\bf 5^{\ast}}=  - \frac{1}{32 \pi^2}
(F_{\mu \nu}^{a} \tilde{F}_{\mu \nu}^{a} )  
\ \ ,
\ee
\be
\label{Mozart2}
J^{\mu}_{\bf 10}= \bar{\chi}^{[ij]}\bar{\sigma}^{\mu}\chi_{[ij]}
\ \ , \ 
\de_{\mu} J^{\mu}_{\bf 10}=  -3  \cdot \frac{1}{32 \pi^2}
(F_{\mu \nu}^{a} \tilde{F}_{\mu \nu}^{a} )  
\ \ , 
\ee
as well as the fermion number current
\be
J^{\mu}_{F} = J^{\mu}_{\bf 5^{\ast}}+ J^{\mu}_{\bf 10}
\ \ ,
\ee
whose anomaly is given by
\be
\label{Mozart3}
\de_{\mu} J^{\mu}_{F}=  - 4  \cdot \frac{1}{32 \pi^2}
(F_{\mu \nu}^{a} \tilde{F}_{\mu \nu}^{a} )  
\ \ .
\ee
From  (\ref{Mozart3}) it follows that the $U(1)$ group 
associated with fermion number (non)conservation, 
\be
\label{u1r}
U(1): \ \ \chi^{i}\longrightarrow e^{i\alpha} \chi^{i}\ , \ \ 
\chi_{[ij]} \longrightarrow e^{i\alpha} \chi_{[ij]} \ \ , 
\ee
has a discrete $\zet_{4}$ subgroup
generated by the transformations 
(\ref{u1r}) with $\alpha_{m}=(2\pi / 4 ) m$, $m\in \zet $
which is a symmetry of the full quantum theory, since in this case 
the action functional $S$ transforms as 
\be
\label{ciccillo}
S\longrightarrow S + i4 k  \alpha_m= S + 2\pi i m \ \ .
\ee
The $SU(5)$ Wilson--Majorana  term  reads, in continuum notation,
\be
S_{W}[{\bf 5^{\ast}}, {\bf 10}] =-\frac{a}{2} \int\! d^{4}x\ \Bigl( 
r_{ij} \chi^{i}\Box\chi^{j} + 
r^{[ij],[lm]} \chi_{[ij]}\Box\chi_{[lm]}\Bigr)
+ {\rm h. c. } \ \ ,
\ee
where 
$r_{ij}$   ($r^{[ij],[lm]}$) is an arbitrary,  nonsingular matrix,
symmetrical in $i$ and $j$ (in $[ij]$ and $[lm]$).
It is then immediately seen that $S_{W}[{\bf 5^{\ast}}, {\bf 10}]$
possess a $\zet_2$ symmetry for each of the Weyl left--handed fermion.
 
Of course, one could wonder if this discrete symmetry is spontaneously 
broken in the infinite volume limit. 
In order to fully understand the question  of 
fermion number violation and to clarify 
in which way the $\zet_2$ is realized,  
one should check that the model
has the correct anomaly structure for the left--handed global current.
To investigate this important issue, we plan to compute 
the abelian axial anomaly with Wilson--Majorana fermions. 

In the following section we shall study the perturbative behavior of
the model. We shall see that it correctly reproduce 
the continuum perturbation theory properties.

\section{Perturbative Renormalization and \newline
One--Loop Calculations} 
\setcounter{equation}{0}
The consequence of introducing   the Wilson--Majorana regularization is 
that the Slavnov--Taylor identities corresponding to the would--be
\mbox{BRST} symmetry do not hold when \mbox{$a \neq 0$}. This allows 
an {\it a priori}
 large variety of nonsymmetrical contributions to Feynman
amplitudes: the question of 
the possible recovering of the chiral symmetry then arises.
The Green functions of the theory defined by \mbox{$S + S_{c.t.}$} 
can satisfy the \mbox{BRST} identities of 
the formal continuum theory only 
up to terms vanishing when  \mbox{$a \rightarrow 0$}.
In this picture, the only purpose of  the formal bare theory (the ``target 
theory" of \cite{Borrelli}) is to provide us with 
a number of equations sufficient to 
determine the coefficients of the monomials in \mbox{$S_{c.t.}$}. Then it will 
be possible to perform  the final renormalization, 
\ie\ to match  the (new) 
theory  \mbox{$S + S_{c.t.}$} and the continuous one.
Equivalently, we could  firstly  perform a lattice 
renormalization with the only 
purpose of fixing the unknown coefficients contained in \mbox{$S_{c.t.}$} 
by requiring that the \mbox{$a \rightarrow 0$} limit of the Green  functions 
of \mbox{$S + S_{c.t.}$}  coincides with the  continuum  result (as a 
consequence they will also obey the same \mbox{BRST} identities).

The possibility of restoring gauge invariance in  a chiral gauge theory
adding 
 a suitable collection of  nongauge--invariant, non Lorentz--invariant 
(on the  lattice) and also 
non \mbox{$U(1)$}--invariant (in the Wilson--Majorana 
version  discussed here) terms is suggested by the 
circumstance that we can consider a  noninvariant regularization  of 
 \mbox{QCD} (or \mbox{QED}) 
even though we know how to \mbox{gauge--}invariantly regularize vectorial 
theories. The difference with the present situation lies in the fact that the 
gauge--invariant regularization satisfies the   \mbox{Slavnov--Taylor} 
identities, making it possible, through the fine tuning of the free parameters 
in \mbox{$S_{c.t.}$}, to obtain the  invariant continuum theory 
even starting from 
the  noninvariant 
regularization, at least in perturbation theory.
In a chiral gauge theory the regularized Green functions 
\mbox{$G_{\Lambda}(x_1 , \ldots , x_n)$} are {\it a priori}
 unconstrained, 
thus casting doubts on the feasibility  of  this program.
We stress that this fine tuning must be performed  in a nonperturbative 
way \cite{Parisi}, particularly for the coefficients of the dimensionful 
couplings.  However the perturbative 
way allows us to perform some simple analytical calculations 
and checks (at least to  lowest order), that 
can be useful to study renormalizability  and to confirm that
anomalous dimensions are as in any other continuum regularization.
\subsection{Vertex Function and Propagators}
Here we discuss the structure of the  Feynman diagrams  we have
calculated perturbatively starting from the action functional
$S$ defined in \mbox{(\ref{action})}. They are the vacuum polarization
tensor, the vertex function and the fermion \mbox{self--}energies (of course 
we have three different  \mbox{self--}energies, as in this theory
there are three different propagators for our Weyl fields 
$\chi^{\alpha}$, $\bar{\chi}_{\dot{\alpha}}$).
We shall then see how these ingredients can be used to fix a 
number of equations for the unknown coefficients of the 
interactions contained in \mbox{$S_{c.t.}$}.

\subsubsection{Vacuum Polarization}
We begin by considering the vacuum polarization tensor
\mbox{$\Pi_{\mu \nu}^{ab}(k)$}, defined by\footnote{We work in the Feynman 
gauge, $\alpha=1$.}
\be
\int\! d^{4}x \ e^{-ikx}
\left\langle A_{\mu}^{a}(x) A_{\nu}^{b}(0) \right\rangle \ = \  
\frac{\delta_{\mu \nu}\delta^{ab}}{k^2}\ + \ 
\frac{\delta_{\mu \rho}}{k^2}\Pi_{\rho \sigma}^{ab}(k) 
\frac{\delta_{\sigma \nu}}{k^2} \ + \ \cdots
\ee
We find that
\bea
\label{pimunu}
\Pi_{\mu \nu}^{ab}(k)\ = \ 
g^2 \ \delta_{\mu \nu} \delta^{ab} \Bigl[ \delta M^{2}_{a}(r)  + 
k^2 G_{1}^{a}(r) +  k_{\mu}^{2} G_{n.c.}^{a}(r) \Bigr] \ +  
\nonumber  \\ 
g^2 \ \delta^{ab}\Bigl( k^2 \delta_{\mu \nu} - k_{\mu} k_{\nu} \Bigr) 
\Bigl[ G_{0} \ln a^2 k^2  \ + \ G_{F} (r)\Bigr]     \ \ .
\eea
Here the \mbox{$U(1)$} symmetry breaking shows up  very clearly:
we can not only make fermion--antifermion pairs 
circulate in loops, but also 
fermion--fermion as well as antifermion--antifermion pairs, since we do not 
have to respect fermion number conservation. The consequence on the 
vacuum polarization tensor is the explicit dependence of the coefficients
\mbox{$\delta M^{2}_{a}(r)$}, \mbox{$G_{1}^{a}(r)$}, \mbox{$G_{n.c.}^{a}(r) $}
in (\ref{pimunu}) on the color \mbox{index $a$}.
Note that the only logarithmic divergence is the same of  the continuum one. 
Moreover, using dimensional regularization  we see  that the coefficient of 
this divergence is again
equal to \mbox{$G_{0}$}, where  $G_0 = 1/48\pi^2$. 
 On the other hand, a mass term 
\mbox{$\delta M^{2}_{a}(r)\ \propto \ r a^{-2}$} has emerged, 
which renormalize differently
the mass of the gauge fields $A_{\mu}^{a}$ 
for different values of the color index   $a=1,2,3 $.
Of course, in a vectorial theory, like \mbox{QCD}  or  \mbox{QED}, 
regularized in a \mbox{gauge--}invariant way (\mbox{\it e.g.}, 
on the lattice, by 
suitably connecting the fermionic matter fields in the Wilson term with the 
gauge links), a term like this can not appear. In our case, however, 
\mbox{gauge} invariance (or, with the \mbox{gauge--}fixed action,
\mbox{BRST} invariance) is broken by the \mbox{Wilson--}Majorana term, and we 
can only expect that it has to vanish when \mbox{$r \rightarrow 0$}, as we have 
verified.
The same arguments apply to the coefficient \mbox{$G^{a}_{n.c.}(r)$} of the 
\mbox{$k_{\mu}^{2}\delta_{\mu \nu}$} term, which explicitly breaks 
\mbox{Lorentz} invariance (still satisfying the 
unbroken hypercubical lattice symmetry), 
as well as to \mbox{$G_{1}^{a}(r)$}.

Let us briefly discuss the structure of the tensor 
$\Pi_{\mu \nu}^{ab}(k)$ in color space.
The regularized action does not display the global \mbox{$SU(2)$} and 
\mbox{$U(1)$} symmetries of the continuum  theory, and it is not 
{\it a priori}
obvious that  it should be  proportional to 
\mbox{$\delta^{ab}$}.  In fact, 
this simple structure stems from the three surviving 
symmetries of the regularized action (that we 
mentioned in the previous section), 
which require 
\[
\begin{array}{c c}
\left\langle A_{\mu}^{a}(x) A_{\nu}^{b}(0) \right\rangle \ = \ 0 \ \ 
\mbox{{\rm if}}  \
\  a \neq  b \ \ \ , & \ \ \ 
\left\langle A_{\mu}^{1}(x) A_{\nu}^{1}(0) \right\rangle \ = \
\left\langle A_{\mu}^{3}(x) A_{\nu}^{3}(0) \right\rangle \  \ .
\end{array}
\]
As a consequence, the general form  of \mbox{$\Pi_{\mu \nu}^{ab}(k)$}
will be 
\be
\label{quattrotre}
\begin{array}{c c}
\Pi_{\mu \nu}^{ab}(k) \ = \ 
\delta^{ab}\Pi_{\mu \nu}^{a}(k) \ \ \ , & \ \ \ 
\Pi_{\mu \nu}^{1}(k) \ = \  \Pi_{\mu \nu}^{3}(k) \ \ .
\end{array}
\ee
We have also checked that the coefficients in  (\ref{pimunu}) 
satisfy  the second condition in (\ref{quattrotre}).
  
\subsubsection{Fermion Self--Energies}
We start with the diagonal propagator of Weyl fields 
\mbox{$S^{\dot{\alpha} \alpha}_{AB}(p)$}, which  
we write as 
\be
\int\! d^{4}x  \ e^{-ipx}
\left\langle \chi^{\alpha}_{A}(x) \bar{\chi}^{\dot{\alpha}}_{B}(0) 
\right\rangle
\ = P^{\dot{\alpha} \alpha}_{AB}(p)  \ + \ 
P^{\dot{\alpha_1} \alpha}_{AC}(p)\biggl[
i\ \Sigma_{\dot{\alpha}_1 \beta_2} (p) \biggr]_{CD}
P^{\dot{\alpha} \beta_2}_{DB}(p) \ + \ \cdots \ ,
\ee
where
\be
\Bigl[ \Sigma_{\dot{\alpha} \beta}\Bigr]_{AB}(p)= 
g^2 p_{\dot{\alpha} \beta}\ 
\delta_{AB}\ \Bigl[ A \ \log a^2 p^2 \ + \  B(r)\Bigr]  \ .
\ee
As in  (\ref{pimunu}), dimensional regularization 
gives the same result for the coefficient 
of the logarithmic pole,
$A=C_F / 16\pi^2$, where $C_F = T^{b}_{f}T^{b}_{f}$.
For $SU(2)$, $C_F=3/4$.

We can observe that the complete propagator is still diagonal in color indices; 
this can be explained using the discrete symmetries 
(\ref{d2}), (\ref{d3})
of the regularized action, 
which imply 
\[
\left\langle \chi^{\alpha}_{A}(x) \bar{\chi}^{\dot{\alpha}}_{B}(0) \right\rangle
\propto  \ \delta_{AB} \ \ . 
\]
In the following, we shall need to extract the \mbox{$r\rightarrow 0^{+}$} 
part of the Feynman amplitudes, which are automatically \mbox{BRST} invariant,
as in \cite{Borrelli}: then we write
\be
\Sigma_{\dot{\alpha} \beta}(p)= 
\Bigl[ \Sigma_{\dot{\alpha} \beta}(p) \Bigr]_{r=0^{+}}\ + \ 
g^2 p_{\dot{\alpha} \beta} G(r) \ , 
\ee
where now  \mbox{$G(0) = 0$}. 

As remarked before,  the global
\mbox{$U(1)$} symmetry which would ensure fermion number conservation,
 is broken at the 
cutoff level by the  \mbox{Wilson--}Majorana term, 
which formally vanishes when
\mbox{$a \rightarrow 0$}. Consequently, 
the  zeroth order approximation, 
\mbox{$P^{\alpha\beta}_{AB}(p) $}, of the nondiagonal 
propagator,  \mbox{$S^{\alpha\beta}_{AB}(p) $}, also vanishes in this 
limit. However, when inserted in loops, it can give rise 
to finite as well as to 
ultraviolet divergent contributions to the Feynman amplitudes: 
in the present case it produces a Majorana mass  renormalization, 
which  for finite $a$ reads 
\be
\int\! d^{4}x  \ e^{-ipx}
\left\langle \chi^{\alpha}_{A}(x) \chi^{\beta}_{B}(0) \right\rangle \ =  
P^{\alpha\beta}_{AB}(p) +
P^{\dot{\alpha}_1 \alpha}(p)\Bigl[
\Sigma_{\dot{\alpha}_1\dot{\alpha}_2}\Bigr]_{AB}
P^{\dot{\alpha}_2 \beta}(p) \ + \ \cdots \ ,
\ee
where 
\be
\label{quattrootto}
\Bigl[ \Sigma_{\dot{\alpha}\dot{\beta}}\Bigr]_{AB}=
g^2 \delta_{AB} \ \epsilon_{\dot{\alpha} \dot{\beta}} \ \delta m (r) \ ,
\ee
with \mbox{$\delta m(r) \ \propto \ r a^{-1}$} and, 
as for the bosonic mass term,
\mbox{$\delta m(0) = 0$}. When the cutoff is removed, the two diagonal 
propagators in the second term tend  to their continuum limit, 
\mbox{$P^{\dot{\alpha} \beta}(p) \rightarrow 
-i p^{\dot{\alpha} \beta} / p^2$}, 
and the result is a linearly divergent term.
In passing, we remark that  the diagonal color structure of 
\mbox{$\Bigl[ \Sigma_{\dot{\alpha}\dot{\beta}}\Bigr]_{AB}$} is again a 
direct consequence of the discrete symmetries of the theory
we recorded in the last section.

We omit to write down  the obvious formula for the 
\mbox{self--}energy  of  antimatter fields, which 
is obtained from (\ref{quattrootto}) by
replacing \mbox{$\epsilon_{\dot{\alpha}\dot{\beta}}$} with
\mbox{$\epsilon_{\beta\alpha}$}.

\subsubsection{Vertex Function}
Finally, we present the results of our perturbative calculations for the 
one--particle--irreducible \mbox{(1PI)} vertex function 
\mbox{$\tilde{\Gamma}_{\dot{\rho} \rho, \mu; AB}^{a} (p', p) $}.
We find  that
\bea
&&\tilde{\Gamma}_{\dot{\rho} \rho, \mu; AB}^{a} (p', p) \ = \ 
\biggl[ \tilde{\Gamma}_{\dot{\rho} \rho, \mu; AB}^{a}(p', p) \biggr]_{r=0^{+}} 
\ +  
\nonumber  \\  
&&(-ig\overline{\sigma}_{\dot{\rho}\rho, \mu})\biggl[ \Bigl(1 + 
g^2 \gamma_{0} \ln a^2 \mu^2 + g^2 \Gamma_{1}(r) \Bigr) T^{a}_{AB}  \ 
+ \  g^2 \Gamma_{2}(r) \Bigl(T^{a}\Bigr)^{\ast}_{AB} \biggr] 
\eea
where we have included in 
the first term the \mbox{$r\rightarrow 0^{+}$} finite 
corrections to the  zeroth order vertex  
\mbox{$-ig\overline{\sigma}_{\dot{\rho}\rho, \mu} T^{a}_{AB} $}; 
the functions 
\mbox{$\Gamma_{i}(r),  \ i=1,2$} are both
independent of \mbox{$p'  , p$}, and satisfy the condition
\mbox{$\Gamma_{i}(0)\ = \ 0 \ $}.

Let us note, in the previous formula, the appearance of  the 
generators of the color symmetry 
in the antifundamental representation,
\mbox{$\Bigl(T^{a}\Bigr)^{\ast}_{AB} $}: this is due 
to  the insertion of nondiagonal Weyl propagators 
in virtual loops, as an effect of the breaking of global 
\mbox{$SU(2)$} and \mbox{$U(1)$} symmetries in the regularized theory.
Lastly, we have checked that the coefficient of the logarithmic divergence
is again the same as in the continuum, \ie\
$\gamma_0=(C_F + C_{adj}) / 16 \pi^2$, 
$C_{adj}$ being the Dynkin index of the adjoint representation.
For $SU(2)$,  $C_{adj}=2$.

\subsection{Slavnov--Taylor Identities}
Following the ideas sketched in the last section, we now examine a number of 
Slavnov--Taylor identities to determine the structure of \mbox{$S_{c.t.}$}.
We start by considering the gauge field sector. 
The relevant identity is\footnote{We use continuum notation, 
for the sake of simplicity.}
\be
\label{prima}
\left\langle \delta\Bigl(  \bar{c}^{a}(x)\ A_{\nu}^{b}(y) \Bigr) 
\right\rangle \ = \ 0 \  \ ,
\ee
which, using  (\ref{trasfbrst}), becomes 
\be
\label{long}
\frac{1}{\alpha}  \left\langle \partial_{\mu}^{x}  A_{\mu}^{a}(x) 
A_{\nu}^{b}(y) \right\rangle 
\ + \   \left\langle \bar{c}^{a}(x) \Bigl( D_{\nu}  c \Bigr)^{b}(y) \
\right\rangle \  = \ 0   \ \ .
\ee
Differentiating with respect to \mbox{$y_{\nu}$} and 
using the equations of motion for ghost fields, we get 
\be
\partial_{\mu}^{x} \partial_{\nu}^{y}
\left\langle   A_{\mu}^{a}(x) \ A_{\nu}^{b}(y) \right\rangle  =  
\alpha  \ \delta^{(4)}(x-y) \ \ ,  
\ee
which shows that the longitudinal part of the gauge field propagator is 
unrenormalized  by the interaction.
This identity is automatically satisfied by the pure gauge and ghost 
contributions to the vacuum polarization tensor even  for \mbox{$a \neq 0$} 
(if we \mbox{gauge--}invariantly regularize this sector), but {\em not} by 
terms arising from  fermion matter fields. 
By adding  to the bare  action $S$  the  appropriate  
counterterms of dimension less than or equal to four we called  
\mbox{$S_{c.t.}$}, we will have
\[
\Pi_{\mu \nu}^{ab} (k) \longrightarrow 
\Pi_{\mu \nu}^{ab} (k) \ + \ \Bigl[ \Pi_{\mu \nu}^{ab}\Bigr]_{c.t.}(k) \ \ .
\] 
Is it now possible to satisfy the identity (\ref{prima})
up to terms of \mbox{O($a$)} ? 
By Fourier transforming into momentum space we obtain the 
following equation
\be
\label{quattrotredici}
k_{\mu} k_{\nu} \biggl( \Pi_{\mu \nu}^{ab} (k) \ + \ 
\Bigl[ \Pi_{\mu \nu}^{ab}\Bigr]_{c.t.}(k) \biggr)
\ = \ 0 \ \ ,
\ee
which fixes a number of coefficients for the interactions in \mbox{$S_{c.t.}$}.
In fact, (\ref{quattrotredici})  requires 
\be
\Bigl[ \Pi_{\mu \nu}^{ab}\Bigr]_{c.t.}(k) = 
- \ g^2 \ \delta^{ab} \Bigl[ \delta_{\mu \nu} \delta M^{2}_{a}(r)  
+ k_{\mu}^{2}\delta_{\mu \nu} G_{n.c.}^{a}(r) + 
k_{\mu} k_{\nu} G_{1}^{a}(r)\Bigr]  \  \ .
\ee
This means that we have to add to the bare action a 
gauge-- and \mbox{Lorentz--}variant (but still 
renormalizable)  counterterm of the form 
\be
\label{c1}
\frac{1}{2}  \int\! d^{4}x \ \sum_{a} \biggl(  \delta M^{2}_{a} 
A_{\mu}^{a} A_{\mu}^{a}  \ + \  G_{n.c.}^{a}(r) \sum_{\mu} 
(\partial_{\mu}A_{\mu}^{a}) (\partial_{\mu}A_{\mu}^{a}) \ + \ 
G_{1}^{a}(r)  (\partial_{\mu}A_{\mu}^{a})^{2}(x) \biggr) \ .
\ee
 
In the fermion sector, we have to impose the 
following  three \mbox{Slavnov--}Taylor identities
to force the regularized theory to reproduce  the continuum theory 
\mbox{BRST} symmetry:
\be
\left\langle \delta\Bigl(  \chi^{\alpha}_{A}(x) 
\bar{\chi}^{\dot{\beta}}_{B}(y) 
\bar{c}^{a}(z)\  \Bigr) 
\right\rangle \ = \ 0 \  \ ,
\ee
\be
\begin{array}{c c}
\left\langle \delta\Bigl(  \chi^{\alpha}_{A}(x) \chi^{\beta}_{B}(y) 
\bar{c}^{a}(z)\  \Bigr) 
\right\rangle \ = \ 0 \  \ \ , & \ \ \ 
\left\langle \delta\Bigl(  \bar{\chi}^{\dot{\alpha}}_{A}(x) 
\bar{\chi}^{\dot{\beta}}_{B}(y) 
\bar{c}^{a}(z)\  \Bigr) 
\right\rangle \ = \ 0 \ \  ,
\end{array}
\ee
that is
\bea
\label{brst1f}
&&\frac{1}{\alpha}  \left\langle \partial_{\mu}^{z}  A_{\mu}^{a}(z) 
\chi^{\alpha}_{A}(x) \bar{\chi}^{\dot{\beta}}_{B}(y)\right\rangle 
 -   ig  \left\langle \bar{c}^{a}(z) c^{b}(x) 
\Bigl(T^b   \chi^{\alpha} (x) \Bigr)_{A} \bar{\chi}^{\dot{\beta}}_{B}(y) 
\right\rangle 
\nonumber \\  
&& +   
ig  \left\langle \bar{c}^{a}(z) 
\chi^{\alpha}_{A} (x) \Bigl( \bar{\chi}^{\dot{\beta}} (y) T^b \Bigr)_{B} 
c^{b}(y)
\right\rangle  =  0    \ \ ,
\eea
\bea
\label{brst2f}
&&\frac{1}{\alpha}  \left\langle \partial_{\mu}^{z}  A_{\mu}^{a}(z) 
\chi^{\alpha}_{A}(x) \chi^{\beta}_{B}(y)\right\rangle 
 -  
ig  \left\langle \bar{c}^{a}(z) c^{b}(x) 
\Bigl(T^b   \chi^{\alpha} (x) \Bigr)_{A} \chi^{\beta}_{B}(y) 
\right\rangle 
\nonumber \\  
&& +  
ig  \left\langle \bar{c}^{a}(z) 
\chi^{\alpha}_{A} (x) c^{b}(y)  \Bigl( T^b \chi^{\beta} (y) \Bigr)_{B} 
\right\rangle \  = \ 0   \ \ ,
\eea
\bea
\label{brst3f}
&&\frac{1}{\alpha}  \left\langle \partial_{\mu}^{z}  A_{\mu}^{a}(z) 
\bar{\chi}^{\dot{\alpha}}_{A}(x) \bar{\chi}^{\dot{\beta}}_{B}(y)
\right\rangle 
 + 
  \ ig  \left\langle \bar{c}^{a}(z)  c^{b}(x)
\Bigl(\bar{\chi}^{\dot{\alpha}}(x) T^b    \Bigr)_{A} 
\bar{\chi}^{\dot{\beta}}_{B}(y) \right\rangle 
\nonumber \\  
&&-
ig  \left\langle \bar{c}^{a}(z) 
\bar{\chi}^{\dot{\alpha}}_{A} (x)   c^{b}(y)
\Bigl( \bar{\chi}^{\dot{\beta}}(y) T^b  \Bigr)_{B}  \right\rangle \  = \ 0   
\ \ .
\eea
We start by examining  the first identity. 
Writing  the  fermion--fermion--gauge  Green function in terms of 
\mbox{1PI} vertex function, complete fermion propagators and complete
gauge field propagator, whose longitudinal part is fixed by means of 
(\ref{long}), we get
\bea
&&\frac{1}{\alpha}  \int\! d^{4}x d^{4}y \  e^{iqx + ipy }  
\left\langle \partial_{\mu}^{x}  A_{\mu}^{a}(x) 
\chi^{\alpha}_{A}(0) \bar{\chi}^{\dot{\beta}}_{B}(y)\right\rangle \ = 
\nonumber \\ 
&&-i \ \frac{q_{\nu}}{q^2}\ S^{\dot{\rho} \alpha}(q+p) 
\tilde{\Gamma}_{\dot{\rho} \rho, \nu; AB}^{a} (q+p, p) \ S^{\dot{\beta} \rho}(p)
\ .
\eea
The second and the third term of the identity (\ref{brst1f})
are seen to be, for 
\mbox{$a\neq 0$}, 
\bea
\label{kappa2}
\sum_{x,y} a^8 \   e^{iqx + ipy }  
\left\langle \bar{c}^{a}(x) c^{b}(0) 
\Bigl(T^b   \chi^{\alpha} (0) \Bigr)_{A} \bar{\chi}^{\dot{\beta}}_{B}(y) 
\right\rangle \ = 
\nonumber \\
- \triangle_{gh}(q) T^{a}_{AB} \Bigl[ S^{\dot{\beta} \alpha} (p) \ + \ 
\overline{\sigma}^{\dot{\rho} \alpha}_{\tau} \overline{\sigma}_{\dot{\rho} 
\rho , \lambda} S^{\dot{\beta} \rho} (p) \ g^2 K_{\tau \lambda}(p, q) \Bigr] 
\ \ ,
\eea
\bea
\label{kappa3}
&&\sum_{x,y} a^8 \   e^{iqx + ipy }  
\left\langle \bar{c}^{a}(x) \chi^{\alpha}_{A} (0) 
\Bigl( \bar{\chi}^{\dot{\beta}} (y) T^b \Bigr)_{B} c^{b}(y)
\right\rangle \  = 
\nonumber \\
&& - \triangle_{gh}(q) T^{a}_{AB} 
\Bigl[ S^{\dot{\beta} \alpha} (p + q)  \ + \
\nonumber \\
&&
S^{\dot{\rho} \alpha} (p + q) \overline{\sigma}_{\dot{\rho} \rho ,\lambda} 
\overline{\sigma}^{\dot{\beta} \rho}_{\tau} \ 
g^2 K_{\tau \lambda}(-p -q , q)  \Bigr]  \ \ ,
\eea
where \mbox{$\triangle_{gh}(q)$} is the ghost propagator.
\mbox{$K_{\tau \lambda}(p, q) $}  is a complicated
function of the external momenta \mbox{$p,q$} and of the Wilson parameter $r$;
its complete expression can be found in Appendix B.
The crucial point is that, extracting the  \mbox{$r\rightarrow 0^{+}$} part, 
we obtain 
\be
\label{kappa}
K_{\tau \lambda}(p, q) \ = \ \Bigl[ K_{\tau \lambda}(p, q) \Bigr]_{r=0^{+}}
\ + \  \delta_{\tau \lambda}\ K(r) \ \ ,
\ee
where now \mbox{$K(r)$} {\em only} depends  on $r$.
Substituting these expressions into    (\ref{brst1f}), the 
\mbox{$r=0^{+}$}  terms in (\ref{kappa})
satisfy this identity,  whereas the $r$--dependent
ones do not.  We are thus lead 
to introduce a new interaction in the 
bare action in order to compensate 
for these terms, \ie\
\be
\label{cing}
Z_{2}\int\! d^{4}x \ \bar{\chi}^{\dot{\alpha}}_{A} 
\partial_{\dot{\alpha} \beta , \mu}
\chi^{\beta}_{A}  \ + \ 
Z_{1}\int\! d^{4}x \ \bar{\chi}^{\dot{\alpha}}_{A} (ig A_{\dot{\alpha}\beta})
_{AB} \chi^{\beta}_{B}  \ + \ 
\tilde{Z}_{1}\int\! d^{4}x \ \bar{\chi}^{\dot{\alpha}}_{A}
(ig A_{\dot{\alpha}\beta})_{BA}\chi^{\beta}_{B} \ .
\ee
Imposing   (\ref{brst1f}), we get the following
equations:
\be
\begin{array}{c c}
Z_{1}-Z_{2}\ = \ - \ g^2  \Bigl(  G  +  \Gamma_{1} + K \Bigr) (r)\ \  \ \  , 
& \ \ \tilde{Z_{1}}\ = \ -\ g^2 \Gamma_{2}(r)\ \ .
\end{array}
\ee
It will not be possible (and actually it is not necessary) to fix 
separately the two constants  \mbox{$Z_{1} , Z_{2}$}. In fact, 
we are only 
demanding that, for a generic string of fields 
\mbox{$O(x_1 , \ldots , x_{n})$}, the identities
\[
\left\langle \delta O (x_1 , \ldots , x_{n}) \right\rangle \ = \ 
\left\langle O (x_1 , \ldots ,  x_{n}) \ 
\delta  (S_{W} + S_{c.t.}) \right\rangle \ = \  0 
\]
should be 
satisfied up to terms vanishing when  \mbox{$a \rightarrow 0$}.  
When the counterterm  (\ref{cing}) is inserted 
in the previous equation, only 
the difference  \mbox{$Z_{1} - Z_{2}$} can appear,   since 
\be
\delta    \biggl( \bar{\chi}^{\dot{\alpha}}_{A} 
(\partial_{\dot{\alpha} \beta} + 
ig A_{\dot{\alpha}\beta})_{AB} \chi^{\beta}_{B}  \biggr) \ = \ 0 \  \ .
\ee
Of course, when the final renormalization is performed, all the coefficients of 
the counterterms will acquire their \mbox{well--}definite values, 
once appropriate  renormalization conditions are imposed.

Finally, we briefly examine identities (\ref{brst2f}) and (\ref{brst3f}). 
In  the one loop approximation, non \mbox{BRST--}invariant contributions  
can only arise from first  order corrections to the 
nondiagonal propagators for fermion fields, which  
are responsible for  a Majorana 
mass renormalization. The introduction of a  counterterm of the form 
\be
\label{cm}
\frac{M}{2}\int\! d^{4}x \ 
\Bigl( \bar{\chi}^{\dot{\alpha}}\bar{\chi}^{\dot{\beta}}
\epsilon_{\dot{\alpha} \dot{\beta}} + \chi^{\alpha} \chi^{\beta} 
\epsilon_{\beta \alpha} \Bigr)
\ee
will make the corresponding identity satisfied, provided that 
\be
M \ = \ \delta m (r) \ \ .
\ee
The same equation could have also been obtained 
by noting that a 
mass  term is not even 
invariant with respect to global \mbox{$SU(2)$} 
or \mbox{$U(1)$}
transformations. One of the consequences of this observation
is that we have, for  the two point 
function \mbox{$\left\langle \chi^{\alpha}_{A}(x) \chi^{\beta}_{B}(0) 
\right\rangle$}, 
\be
T^{a}_{AC} \left\langle \chi^{\alpha}_{C}(x) \chi^{\beta}_{B}(0) 
\right\rangle \ + \ T^{a}_{BC}
\left\langle \chi^{\alpha}_{A}(x) \chi^{\beta}_{C}(0) 
\right\rangle \ = \ 0 \ \ ,
\ee
that is, for \mbox{$A=B$},
\be
T^{a}_{AC}\Bigl(  \left\langle \chi^{\alpha}_{C}(x) \chi^{\beta}_{A}(0) 
\right\rangle \ - \ \left\langle \chi^{\beta}_{C}(0) \chi^{\alpha}_{A}(x) 
\right\rangle \Bigr) \ = \ 0 \  \ .
\ee
This  condition  together with a similar one for  antimatter fields  is 
trivially  satisfied just when the counterterm in  
(\ref{cm}) is included 
in the action  functional of the regularized theory.
\section{Conclusions}
\setcounter{equation}{0}
We have presented here a few one--loop calculations in
a simple anomaly--free model of lattice chiral gauge theory. 
They lend strong support to 
the idea that it is possible to reach   
the BRST--invariant, continuum theory, starting 
from an action functional which is the sum of a Wilson regularization 
of the tree level action plus a finite number of renormalizable 
counterterms. In the one--loop approximation, they are
given by (\ref{c1}), (\ref{cing}), (\ref{cm}).
The coefficients of these interactions must be fine--tuned 
in order to match the continuum target theory onto its regularized form. 
Our regularization makes use of the physical, left--handed
degrees of freedom of the theory only, and
in perturbation theory we found no obstructions 
which would prevent this program to be carried out successfully.
The correctness of this procedure at the nonperturbative level
cannot be settled at this stage of work, and should be assumed in any
forthcoming Monte Carlo simulation.
As discussed in section 4,
a study of nonperturbative (instanton) physics could clarify the issue of
the existence of a $\zet_2$ simmetry in the regularized theory.
Since the model  correctly reproduce perturbation theory, 
we do not expect essential problems. 
We plan to come back on this point in a future publication.

\vskip 1.5cm

\leftline{\bf\large Acknowledgments}
\nopagebreak
\vskip .4cm
\nopagebreak
It is a great pleasure to thank my thesis advisor, Massimo Testa, for 
suggesting  me this problem,  for his constant encouragement and 
support, and for many interesting  discussions.
I would also like to thank 
Francesco Fucito and Gian Carlo Rossi for
a critical reading of this manuscript, and the Referee for
comments and discussions on the issue of fermion number nonconservation.
\newpage
\appendix
\section{Euclidean $\bar{\sigma}$  Matrices}
\setcounter{equation}{0}
It is well known that there is a \mbox{one--to--one} correspondence between
Weyl spinors with one dotted and one undotted index (four
independent components) and ordinary \mbox{4--vectors} \cite{Waerden}, 
which can be written as
\be
p_{\dot{\alpha}\beta} = \sum_{\lambda =1}^{4} 
p_{\lambda} \overline{\sigma}_{\dot{\alpha} \beta , \lambda} \ ,
\ee
where the four \mbox{$2\times 2$} independent matrices 
$\overline{\sigma}_{\lambda}$
are defined through the anticommutation relations they must obey, \ie\
\be
\label{A1}
\overline{\sigma}_{\dot{\alpha}\beta , \mu}
\overline{\sigma}^{\dot{\rho}\beta}_{\nu}+
\overline{\sigma}_{\dot{\alpha}\beta , \nu}
\overline{\sigma}^{\dot{\rho}\beta}_{\mu}
=2 \delta_{\mu \nu}\epsilon^{\dot{\rho}}_{\dot{\alpha}} \ ,
\ee
\be
\label{A2}
\overline{\sigma}_{\dot{\alpha}\beta ,\mu}
\overline{\sigma}^{\dot{\alpha}\rho}_{\nu} 
+ \overline{\sigma}_{\dot{\alpha}\beta ,\nu}
\overline{\sigma}^{\dot{\alpha}\rho}_{\mu} =
2 \delta_{\mu \nu}\epsilon^{\rho}_{ \ \beta}  \ .
\ee
Starting from  (\ref{A1}), (\ref{A2}), it is possible to derive 
some very useful relations for products (and traces of products) 
of  \mbox{$\bar{\sigma}$}
matrices. We collect here some of these properties in 
\mbox{$D$ dimensions}.
\be
\sum_{\mu} \overline{\sigma}_{\mu}^{\dot{\beta} \alpha} 
\overline{\sigma}_{\dot{\rho} \alpha , \mu} = 
D\  \epsilon^{\dot{\beta}}_{\ \dot{\rho}}  \ \ ,
\ee
\be
\sum_{\mu}
\overline{\sigma}_{\dot{\beta} \alpha , \mu}
\overline{\sigma}^{\dot{\gamma} \alpha}_{\rho} 
\overline{\sigma}_{\dot{\gamma} \delta , \mu} \ = \ \left( -D+2 \right) 
\overline{\sigma}_{\dot{\beta} \delta , \rho} 
\ee
\be
\sum_{\mu}
\overline{\sigma}_{\dot{\beta} \alpha , \mu}
\overline{\sigma}^{\dot{\gamma} \alpha}_{\rho_1} 
\overline{\sigma}_{\dot{\gamma} \delta , \rho_2}
\overline{\sigma}^{\dot{\theta} \delta}_{\mu} \ = \ 
4 \ \epsilon_{\rho_1 \rho_2} \epsilon^{\dot{\theta}}_{\ \dot{\beta}}
+ \left( D-4 \right) 
\overline{\sigma}_{\dot{\beta} \delta , \rho_1}
\overline{\sigma}^{\dot{\theta} \delta}_{\rho_2}
\ee
\be
\sum_{\mu}
\overline{\sigma}_{\dot{\beta} \alpha , \mu}
\overline{\sigma}^{\dot{\gamma} \alpha}_{\rho_1} 
\overline{\sigma}_{\dot{\gamma} \delta , \rho_2}
\overline{\sigma}^{\dot{\eta} \delta}_{\rho_3}
\overline{\sigma}_{\dot{\eta} \theta , \mu} \ = \ 
-2\overline{\sigma}_{\dot{\beta} \alpha , \rho_3} 
\overline{\sigma}^{\dot{\gamma} \alpha}_{\rho_2} 
\overline{\sigma}_{\dot{\gamma} \theta , \rho_1}
 - \left( D-4\right)
\overline{\sigma}_{\dot{\beta} \delta , \rho_1} 
\overline{\sigma}^{\dot{\theta} \delta}_{\rho_2} 
\overline{\sigma}_{\dot{\eta} \theta , \rho_3} \ \ ,
\ee
\be
\overline{\sigma}_{\mu}^{\dot{\beta} \alpha} 
\overline{\sigma}_{\dot{\beta} \alpha , \nu} = 
f(D)\  \delta_{\mu \nu} \ \ ,
\ee
where 
\be
f(D) = \mbox{{\rm Tr}} \ \unita  \ \ ;    
\ee
hence $f(4) = 2$.
\section{The Explicit Expression of the Function $K_{\mu \nu} (p,q)$}
\setcounter{equation}{0}
Here we  report the result of our perturbative 
calculations for the  function  $K_{\mu \nu}(p, q)$ 
which appears in  the Slavnov--Taylor identities 
(\ref{kappa2}), (\ref{kappa3}). 
It is readily evaluated to be 
\be
\label{kappa4}
K_{\mu\nu}(p, q)= \frac{C_{adj}}{2} 
\cos \frac{q_{\nu}a}{2} 
\int\! \frac{d^{4}k}{(2 \pi)^{4}} 
\frac{
2 \sin k_{\mu} \sum_{\lambda} \sin \Bigl(\frac{ k + (p+q)a}{2}\Bigr)_{\lambda}
\cos \Bigl( \frac{k-pa}{2}\Bigr)_{\lambda}  \delta_{\lambda \nu}
}{
\Delta (k) \tilde{\Delta} (k + pa) \tilde{\Delta} ( k +  pa + qa ) } \ \ ,
\ee
where
\be
\tilde{\Delta} (k) 
= \frac{1}{4 \sum_{\mu} \sin^{2} \frac{k_{\mu}}{2}} \ \ ,
\ee
and the  function $\Delta (k) $ is defined in (\ref{deftrian}).
It should be pointed out that the only dependence on the Wilson parameter
is contained in the function $\tilde{\Delta}$.
According to the decomposition (\ref{kappa}),  we then obtain from
(\ref{kappa4}), evaluated 
in the limit $a \rightarrow 0$,  the result
\be
K(r) = - \frac{C_{adj}}{8}
\int\! \frac{d^{4}k}{(2 \pi)^{4}} 
\frac{1}{[ \tilde{\Delta} (k) ]^2} 
\left( \frac{D(k)}{\Delta (k)} - 1 \right) 
\ \ ,
\ee
from which it follows that $ K(r) \rightarrow 0$ as
$r \rightarrow 0^{+}$.

\newpage


\bibliographystyle{unsrt}

\end{document}